# Erbium-ytterbium co-doped lithium niobate single-mode microdisk laser with an ultralow threshold of 1 μW

Minghui Li[1,4], Renhong Gao[1,4], Chuntao Li[2,3], Jianglin Guan[2,3], Haisu Zhang[2], Jintian Lin[1,4,8], Guanghui Zhao[1,5], Qian Qiao[1,5], Min Wang[2], Lingling Qiao[1], Li Deng[2], and Ya Cheng[1,2,3, 6,7,*]

[1]State Key Laboratory of High Field Laser Physics and CAS Center for Excellence in Ultra-Intense Laser Science, Shanghai Institute of Optics and Fine Mechanics (SIOM), Chinese Academy of Sciences (CAS), Shanghai 201800, China
[2]XXL—The Extreme Optoelectromechanics Laboratory, School of Physics and Electronic Science, East China Normal University, Shanghai 200241, China
[3]State Key Laboratory of Precision Spectroscopy, East China Normal University, Shanghai 200062, China
[4]Center of Materials Science and Optoelectronics Engineering, University of Chinese Academy of Sciences, Beijing 100049, China
[5]School of Physical Science and Technology, ShanghaiTech University, Shanghai 200031, China
[6]Shanghai Research Center for Quantum Sciences, Shanghai 201315, China
[7]Hefei National Laboratory, Hefei 230088, China
[8]jintianlin@siom.ac.cn
*ya.cheng@siom.ac.cn

We demonstrate single-mode microdisk lasers in the telecom band with ultra-low thresholds on erbium-ytterbium co-doped thin-film lithium niobate (TFLN). The active microdisk were fabricated with high-Q factors by photo-lithography assisted chemo-mechanical etching. Thanks to the erbium-ytterbium co-doping providing high optical gain, the ultra-low loss nanostructuring, and the excitation of high-Q coherent polygon modes which suppresses multi-mode lasing and allows high spatial mode overlap between pump and lasing modes, single-mode laser emission operating at 1530 nm wavelength was observed with an ultra-low threshold, under 980-nm-band optical pump. The threshold was measured as low as 1 μW, which is one order of magnitude smaller than the best results previously reported in single-mode active TFLN microlasers. And the conversion efficiency reaches $4.06 \times 10^{-3}$, which is also the highest value reported in single-mode active TFLN microlasers.

1. Introduction

A high desire for high performance photonic integrated devices such as electro-optic modulators [1-3], optical frequency conversion [4-11], quantum light sources [12-15], optical frequency combs [16-22], and ultra-fast pulse generation [23], rapidly motivates the photonic integration platforms to lithium-niobate-on-insulator (LNOI) wafer [24-27], owing to the outstanding material properties of lithium niobate, such as a broad transparency window (350 nm to 5 μm), a large linear electro-optic, second-order nonlinear, and acoustic-optic, piezo-electric coefficients [24-28]. All the above-mentioned applications have been successfully demonstrated on passive LNOI wafer with unparalleled performances due to the accessible highly confined photonic structures with ultra-low loss due to the rapid developments in ion-slicing technique and LNOI nanofabrication technology. The development of scalable photonic integrated circuits also raises the interest in doping rare-earth ions into LNOI photonic structures to add functionalities enabled by the active ions, for example, microlasers [29-34], optical amplifiers [35-37], and quantum photonic devices [38]. Among the devices, single-mode microlasers have been demonstrated in erbium ion doped microcavities [29-34,39-41], serving as coherent light sources operating in the telecom waveband. The threshold as low as 20 μW was reported [30,42], and the highest conversion efficiency reached $1.8 \times 10^{-3}$ [33]. However, there is still plenty room for improvement in the conversion efficiency and threshold of the single-mode microlasers.

To improve the optical gain of erbium ions, ytterbium ions have been often used as sensitizers to enhance the excitation rates as well as reduce the concentration quenching of erbium ions via resonant energy transfer, facilitating an improved pumping efficiency at the 980-nm band [43,44]. Due to the difficulty in the growth of the co-doped lithium niobate crystal and the fabrication of co-doped thin-film LNOI, erbium-ytterbium ions have only recently been co-doped in LNOI devices, leading to multimode microlasers of low-threshold [45] and optical waveguide amplifiers with a net small-signal gain as high as 27 dB [46]. However, because there are a large number of whispering gallery modes (WGMs) within optical gain bandwidth, single-mode microlasers have not been reported on the co-doped LNOI platform.

In this work, single-mode microlasers was demonstrated in an erbium-ytterbium co-doped LNOI microdisk with a diameter of 40 μm. Weak perturbation was introduced into the circular microdisk to organize the traditional WGMs as polygon modes [30,47]. When the polygon mode in the weakly-perturbed microdisk was excited around 980 nm wavelength, single-mode lasing signal was observed at 1530 nm wavelength, benefiting from the significantly suppression of the large number of WGMs within the gain bandwidth. A laser threshold as low as 1 μW was measured, which is enabled by the high optical gain of co-doping and the excitation of high-Q polygon mode. And the conversion efficiency reaches $4.06 \times 10^{-3}$. Both these values are the-state-of-the-art results in single-mode LNOI microlasers in the telecom band [42].

2. Fabrication methods

The fabrication of erbium-ytterbium co-doped Z-cut LNOI microdisks begins from the growth of the co-doped lithium niobate bulk crystal using by Czochralski method. The doping concentrations of 0.1 mol.% (erbium) and 0.1 mol.% (ytterbium) were adopted to provide high optical gain. Then ion implantation, silica deposition, bonding, chemo-mechanical polishing, and annealing were carried out to produce the co-doped thin-film LNOI wafer. The LNOI wafer consists of 500-nm-thick co-doped lithium niobate thin film, 2-μm-thick silica layer, and 500-μm-thick silicon handle. The microdisks were fabricated on the co-doped LNOI wafer by home-built photo-lithography assisted chemo-mechanical polishing (PLACE) technique [48]. To fabricate the microdisk by PLACE, first, a chromium (Cr) layer with a thickness of 200 nm was deposited on the surface of LNOI by magnetron sputtering method. Subsequently, microdisk-shape pattern was produced in the Cr layer using spatial selective femtosecond laser ablation with a resolution of ~ 200 nm, which serves as hard mask in the next etching step. Next, the chemo-mechanical polishing (CMP) process was conducted to etch the exposed lithium niobate around the Cr mask. Therefore, the pattern was transferred from the Cr layer to the thin-film lithium niobate layer. Third, the sample was immersed in a Cr etching solution for 20 min to remove the Cr layer. Forth, a secondary CMP process was carried out to improve the smoothness of the fabricated LNOI microdisk. Finally, the silica underneath the lithium niobate microdisk was partially undercut by chemical wet etching. The scanning electron microscopy (SEM) images of the fabricated microdisk were shown in Fig. 1, where the diameter of the microdisk was 40 μm.

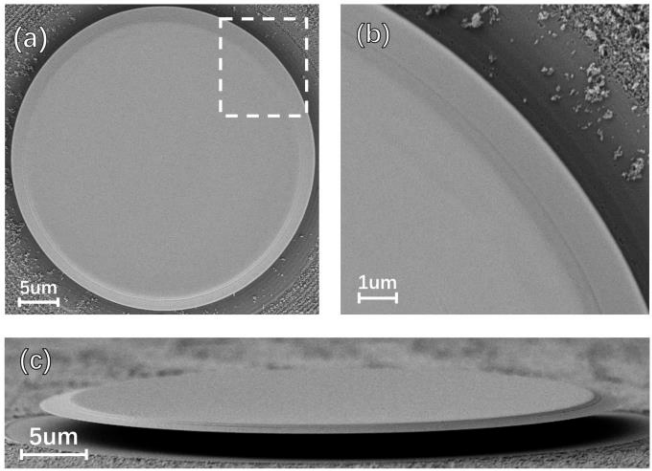

**Fig. 1** The SEM images of the microdisk. (a) Top view. (b) Enlarged SEM image of the microdisk indicated by a rectangle in Fig. (a), showing a smooth side wall. (c) Oblique view.

3. Experimental setup for the microlaser and Q-factor measurement

To demonstrate single-mode microlaser from polygon modes, the Q factors of the pump mode and lasing modes were characterized. The experimental setup for Q-factor measurement and lasing was illustrated in Fig. 2(a). A tunable laser (Model: TLB-6719, New Focus Inc.) with linewidth < 200 kHz connected with an inline polarization controller (PC) was used as pump light, allowing a large wavelength scanning range from 940 nm to 985 nm. The pump light was coupled into the microdisk through a tapered fiber with a waist of 2.0 um. The tapered fiber was placed in close contact with the top surface of the circular microdisk at the position which was ~ 18.5 μm far from the disk center to introduce weak perturbation for the formation of polygon modes resonant with the lasing and pump wavelengths. The coupled position was controlled by a 3D piezo-electric stage with a resolution of 20 nm.

A microscope imaging system consisting of an objective lens with numerical aperture of 0.28 and a charge coupled device was amounted above the weakly perturbed microdisk to monitor and capture the optical field distribution in the microdisk. The generated signal was coupled out of the microdisk by the same tapered fiber, and separately sent to an optical spectrum analyzer (Model: AQ6370D, Yokogawa Inc.) and a photo detector (Model: 1611-FC, New Focus, Inc.) connected with an oscillation (Model: MDO3104, Tektronix Inc.) for optical spectrum analysis and transmission spectrum measurement, respectively. To measure the Q factor of the pump polygon mode, a ramp signal was sent to the tunable laser for fine wavelength scanning across the resonant wavelength. A weak input power as low as 0.5 μW was chosen to avoid thermal-optic effect and lasing. And the transmission spectrum of the tapered fiber was record by the oscillation during wavelength scanning. To measure the Q factor of the lasing polygon mode, another tunable laser (Model: TLB-6728, New Focus Inc.) was replaced with the above-mentioned laser, working in the telecom band. Once the pump wavelength was adjusted to resonant with the pump polygon mode with above-threshold pump power, lasing signals were detected and record by the optical spectrum analyzer (OSA).

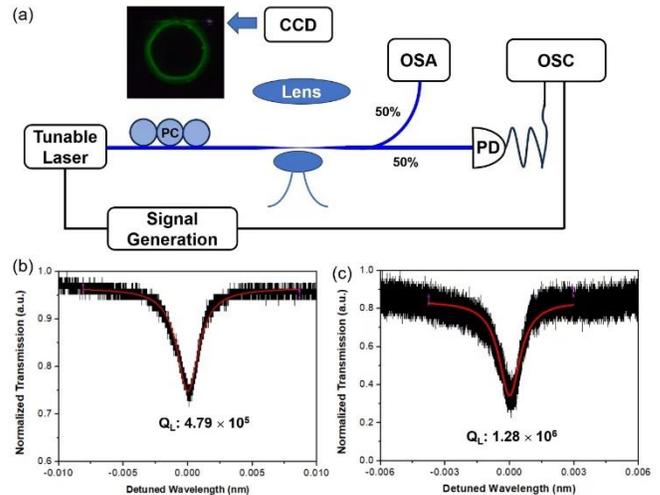

**Fig. 2** (a) Schematic of experimental setup. Inset: the optical microscope image of the up-conversion fluorescence distributed in the microdisk, exhibiting a hexagon pattern. OSA: optical spectrum analyzer. PD: photo detector. PC: polarization controller. CCD: charge coupled device. (b) Lorentz fitting (red curve) of the pump mode around 974.79 nm, revealing a loaded Q factor of 4.79 × $10^5$. (c) Lorentz fitting (red curve) of the lasing mode around 1531.40 nm, exhibiting a loaded Q factor of 1.28 × $10^6$.

Sharp dips will appear in the transmission spectrum when the scanning wavelength is resonant with the modes. The Q factors of the pump and lasing polygon modes are plotted in Figs. 2(b) and (c), respectively. The spatial distribution of the up-conversion fluorescence of the pump light is plotted in the inset of Fig. 2(a), showing a hexagon pattern. Since the up-conversion fluorescence is excited by the pump light, it records the optical field distribution of the pump mode, showing that the pump mode possesses hexagon-pattern spatial distribution [30]. And under polygon mode pumping, only the polygon modes other than the WGMs possess enough gain for lasing [30], since the spatial mode overlap between the polygon lasing modes and the polygon pump mode and is much higher than that of lasing WGMs. The loaded Q factor $Q_L$ of the pump mode at 974.79 nm was measured as 4.79×$10^5$ through Lorentz fitting. And the coupling efficiency of pump light was determined as 22.1%. The loaded Q factor $Q_L$ of the lasing mode at 1531.40 nm was measured to $1.28 \times 10^6$ through Lorentz fitting. And the coupling efficiency of pump light was determined as 40.0%. Therefore, both the polygon pump and lasing modes possess high-Q factors, resulting from the ultrasmooth surface of the fabricated microdisk and the very weak perturbation.

## 4. Demonstration of the single-mode microlaser

A stable single-mode lasing signal was observed at 1531.40 nm when the microdisk was under optical pump at 974.79 nm wavelength with on-chip pump power higher than 1 µW, as shown in Fig. 3. The evolution of the output power of the microlaser under different pump powers is plotted in Fig. 3(b), showing a linear growth. The side mode suppression ratio of the microlaser reaches 23.4 dB when the pump power was 325 µW, which is limited by the available pump power on hand. Figure 3(d) plots the output power of the microlaser as a function of the pump power. The output power of the microlaser linearly grows with the increasing pump power, which agrees well with the nature of lasing. And the threshold was determined to be ~ 1.04 µW, which is one order of magnitude lower than the best results reported in single-mode LNOI microlasers [30,42]. And the conversion efficiency as high as $4.06 \times 10^{-3}$ was measured. Both these two values are the best results reported in single-mode LNOI microlasers, benefiting from the erbium-ytterbium ions co-doping and the use of high-Q polygons which provides high spatial mode overlap between the pump and lasing modes.

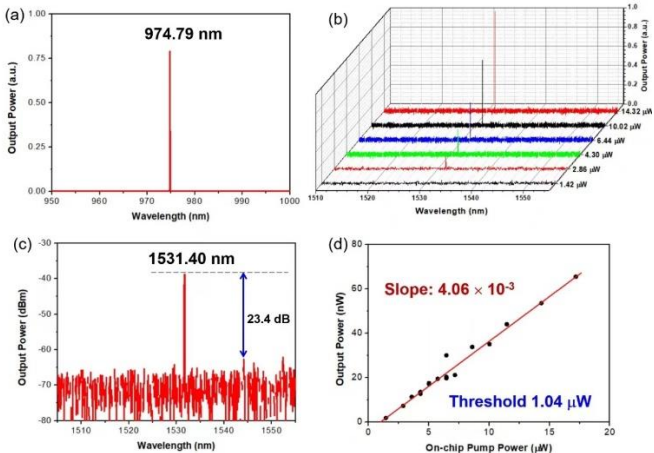

**Fig. 3** (a) The spectrum of the pump light at 974.79 nm wavelength. (b) The evolution of the output power of the microlaser under different pump power levels. (c) The spectrum of lasing signal at 1531.40 nm wavelength, showing a side mode suppression ratio of 23.4 dB. (d) The output power dependence on the pump power, exhibiting an ultra-low threshold of 1.04 µW and a high conversion efficiency of $4.06 \times 10^{-3}$.

## 5. Conclusion

To conclude, we have demonstrated a stable single-mode microdisk laser in the telecom band with an ultralow threshold based on erbium-ytterbium co-doped LNOI chip. Thanks to the erbium-ytterbium co-doping providing higher optical gain, the ultra-low loss nanostructuring by PLACE, and the excitation of high-Q polygon modes allowing high spatial mode overlap and suppression of multimode lasing [30,47,49], a threshold as low as 1 µW was achieved. And a conversion efficiency up to $4.06 \times 10^{-3}$ was reported at room temperature. Such low-threshold single-mode microlaser will promote the development of the scalable photonic integrated circuits on LNOI platform [50-52].